\documentclass[graybox]{svmult}

\usepackage{mathptmx}       
\usepackage{helvet}         
\usepackage{courier}        
\usepackage{type1cm}        
\usepackage{makeidx}         
\usepackage{graphicx}        
\usepackage{multicol}        
\usepackage[bottom]{footmisc}
\usepackage{braket}
\usepackage[colorlinks=true,citecolor=blue]{hyperref}
\usepackage{amsfonts}

\makeindex             

\begin{document}


\title*{Abelian gauge potentials on cubic lattices}

\author{M. Burrello, L. Lepori, S. Paganelli, A. Trombettoni}
\institute{M. Burrello \at Niels Bohr International Academy and Center for Quantum Devices, University of Copenhagen, Juliane Maries Vej 30,
2100 Copenhagen, Denmark, \email{michele.burrello@nbi.ku.dk}
\and
L. Lepori 
\at Dipartimento di Scienze Fisiche e Chimiche, Universit\'a dell'Aquila, via Vetoio,
I-67010 Coppito-L'Aquila, Italy, 
\at INFN, Laboratori Nazionali del Gran Sasso, Via G. Acitelli, 22, I-67100 Assergi (AQ), Italy,\\
\email{llepori81@gmail.com}
\and
S. Paganelli \at Dipartimento di Scienze Fisiche e Chimiche, Universit\'a dell'Aquila, via Vetoio,
I-67010 Coppito-L'Aquila, Italy, \email{pascualox@gmail.com}
\and
A. Trombettoni \at CNR-IOM DEMOCRITOS Simulation Center, Via Bonomea 265, I-34136 Trieste, Italy.
\at SISSA and INFN, Sezione di Trieste,
via Bonomea 265, I-34136 Trieste, Italy, \email{andreatr@sissa.it}
 }

\maketitle

\section{Introduction}

The study of the dynamics of a quantum particle in a magnetic field is 
a fascinating subject of perduring interest both in physics 
and mathematics literature. 
The quantization of energy levels giving rise to the Landau levels 
is at the basis of our understanding of integer and fractional 
quantum Hall effects 
\cite{prange90,bellisard94,yoshioka02,avron03,jain07} 
and its higher-dimensional counterparts and generalizations, 
including topological insulators \cite{bernevig13,ryu2015,prodan16}. 
The use of vector 
potentials in quantum mechanics is associated in itself to very interesting 
consequences, such as the purely quantum mechanical interference in the Aharonov-Bohm effect \cite{aharonov59}. On the other side, the mathematical 
formalism for a particle in a magnetic field has been developed and refined along the time, based on the 
rigorous definition of Schr\"{o}dinger operators with magnetic fields 
\cite{avron78}. An important role is played by the construction 
of families of observables in the presence of gauge fields 
relying on the progresses in 
gauge covariant pseudodifferential calculus \cite{karasev02} 
and the $C^\ast$-algebraic formalism \cite{mantoiu07} (see for a review 
Ref. \cite{mantoiu06} in \cite{asch06}).

A major area of research in the field of single-particle and 
many-body properties in the presence of a magnetic field 
is provided by the study of the effects of periodic potentials.   
The interplay of the magnetic field and the 
discreteness induced by the lattice provides a paradigmatic 
system for the study of incommensurability effects \cite{azbel64,hofstadter76} 
and it results in an energy spectrum 
exhibiting a fractal structure, referred to as the Hofstadter butterfly 
\cite{hofstadter76}. Very interesting examples of the analysis of the so-called colored gaps can be found in 
\cite{avron03_bis,web}, while a discussion of the 
colored Hofstadter butterflies in honeycomb lattices 
can be found in \cite{agazzi14}.  

The study of the Hofstadter Hamiltonian attracted in the years a 
sparkling activity, also due to its connections with the one-dimensional 
Harper model \cite{harper55}. 
A concise, but very clear discussion is presented in 
\cite{thouless90}, where it is shown how the Schr\"odinger equation 
for an electron in a magnetic field in the presence
of a two-dimensional periodic potential can be mapped in a one-dimensional 
quasiperiodic equation. A derivation 
of Harper and Hofstadter models in the context of 
effective models for the conductance in magnetic fields was presented 
in \cite{denittis10}, while a treatment of 
the Schr\"odinger operator in two dimensions with a periodic potential and 
a strong constant magnetic field perturbed by slowly varying 
non-periodic scalar and vector potentials has been recently discussed in 
\cite{freund16}. A remarkable motivation for the studies 
of a two-dimensional electron gas in a uniform magnetic field and 
a periodic substrate potential is as well coming 
from the connection with topological invariants, as one can see 
from the study of the Hall conductance \cite{thouless82}.

The theoretical studies on properties of lattice systems in a periodic 
potential found a experimental matching in the active research of 
solid-state realizations of the Hofstadter and related Hamiltonians. 
This effect has never been observed so far in a natural crystal due to the fact 
that a very large magnetic field would be required, however 
signatures of the Hofstadter bands has been observed 
in artificial superlattices 
\cite{geiser04,melinte04,feil07,dean13,ponomarenko13}. This activity  
found recently a counterpart in the field of cold atoms, where it has 
been possible to load a neutral atomic gas in an optical lattice and simulate 
by external lasers an artificial magnetic potential 
\cite{Aidelsburger2013,Miyake2013}. Given the fact that ultracold atoms, 
due to the high level of control and tunability of parameters 
\cite{stringari16}, are an ideal physical setup in which perform quantum 
simulation \cite{bloch08}, these and related experimental achievements 
opened the way to study a variety of lattice systems in a magnetic potential. 
In particular one can load on the lattice interacting 
bosonic and/or fermionic atoms, 
control the parameters of the lattice, use several components in each 
lattice site \cite{mancini15} and implement a variety of lattices of different 
dimensionalities (not only $D=2$, but also $D=1$ and $D=3$).

The rationale 
of this Chapter is to report an introduction to the different 
ways to implement artificial magnetic potentials in the presence of a controllable 
lattice, with the goal to create a link with the many available results 
in theoretical and mathematical physics. From the other side we think 
that the variety of lattice models in magnetic fields implementable with 
ultracold gases may be a context in which test and apply techniques 
from the mathematical literature, and motivate further analytical and rigorous 
results, starting from the treatment of three-dimensional fermionic lattice systems. With these 
objectives, we then 
present in Section \ref{real} a discussion on several different ways 
of realizing artificial magnetic fluxes in optical lattice systems. In 
Sections \ref{Abelian} and \ref{long_phi} we present two possible 
applications of the results presented in Section \ref{real} both to 
illustrate the versatility of possible uses of artificial magnetic potentials 
and to show results for $3D$ and $1D$ lattices.  
In section \ref{Abelian} we review and study 
cubic lattice tight-binding models with a commensurate Abelian flux, 
also presenting results for the case of anisotropic fluxes. 
In Section \ref{long_phi} we consider $1D$ rings pierced by 
a magnetic field discussing how the latter can enhance 
the quantum state transfer and the entanglement entropy in the system.

We acknowledge several discussions we had along the years on the subjects 
treated in this Chapter with several people, and 
special acknowledgements go 
to M. Aidelsburger, E. Alba, T. Apollaro, H. Buljan, A. Celi,
I. C. Fulga, N. Goldman, G. Gori, M. Mannarelli, G. Mussardo, 
G. Panati, H. Price, M. Rizzi, P. Sodano and A. Smerzi. M. B. acknowledges the Villum Foundation for support. 
S. P. is supported by a Rita Levi-Montalcini fellowship of MIUR.

\section{Realization of artificial magnetic fluxes in optical lattice systems}
\label{real}

In the last ten years many experiments with ultracold atoms 
demonstrated the possibility of realizing artificial magnetic fluxes 
trapped in two-dimensional optical lattices. 
Similar setups pave the way for a systematic study of 
topological phases of matter in the highly controllable environment 
provided by ultracold atoms \cite{bloch08}. Such experiments 
offer, on one side, the possibility of reaching regimes 
that are hardly achievable in solid state devices and, on 
the other, to verify the emergence of topological phenomena through 
observables typical of ultracold gases, 
such as the motion of the center of mass of the system, or the 
momentum distribution of the atoms \cite{stringari16}. 
The measurement of these quantities therefore provides useful tools to 
detect the appearance of topological phases of matter 
which are complementary to the usual transport measurements 
performed in solid state platforms.

The main result which enabled the experimental 
study of topological models in ultracold atomic systems was the 
realization of artificial gauge potentials. In this framework, we speak 
about {\em artificial} (or {\em synthetic}) 
gauge potentials because ultracold atoms are neutral, 
therefore their motion is not directly affected by the presence of a 
true electromagnetic field. Despite that, however, it turned out to be possible 
to engineer systems in which the dynamics of the slow and low-energy 
degrees of freedom can be described by an effective Hamiltonian 
in which the free, non-interacting, part is the one of 
a free-particle in a magnetic field. An interesting point to be observed 
is that the obtained Hamiltonian, featuring the presence of an artificial 
magnetic field, 
is interacting, with the interaction term tunable by e.g. 
Feshbach resonances or by acting on the geometry of the system.  
At the same time, typically, as we are going to discuss in the following, 
the synthetic field does not depend on the interactions or on the density 
of the system, being in a word a single-particle effect.

An efficient way to implement a synthetic gauge potential  $\vec{A}$
giving rise to an artificial, static magnetic field $\vec{B} \propto \nabla\times{\vec{A} }$, 
is to implement with ultracold atoms in optical lattice a tight-binding Hamiltonian with hopping amplitudes which are, in general, complex and whose phases depend on the 
position. To be more clear, we point out that the 
implementation of synthetic gauge potentials in lattices 
relies on the well-established experimental successes in the quantum simulation of tight-binding Hamiltonians for both bosons and fermions \cite{bloch08}.

For ultracold bosons, if a condensate is loaded in an optical lattice, then 
one can expand the condensate 
wavefunctions in the basis of the Wannier functions and obtain a discrete nonlinear Schr\"odinger (DNLS) equation 
\cite{trombettoni01}. The coefficient of the nonlinear term 
in the DNLS equation is proportional to the $s$-wave scattering length $a$ 
and in general the coefficients of the DNLS equation 
depend on integrals of the Wannier functions. 
For $a=0$ (i.e., for an effectively non-interacting condensate) 
one gets the discrete linear Schr\"odinger, which is nothing but 
the tight-biding model for which one can apply 
consolidated numerical analyses \cite{marzari97} and 
rigorous \cite{brouder07} 
techniques for the definition and 
determination of the Wannier functions and their behaviour. 
When $a \neq 0$ a rigorous theory of (nonlinear) Wannier functions does 
not exist, and the semiclassical equations of motion should be modified 
as a consequence of the existence of the nonlinearity. The development 
of a rigorous extension of Wannier functions in the presence of nonlinearity 
is a challenging mathematical problem for the future. We think this 
independently from the fact that an approximate determination 
of the Wannier functions (see e.g. for a variational 
approach in \cite{cataliotti01,vanoosten01,trombettoni05}) 
typically works very well to describe the experimental results 
when the laser intensity, i.e. the strength of the periodic potential, 
is large enough, also when the system approaches the superfluid-Mott 
transition and/or the gas is not longer condensate due to the presence 
of strong interactions \cite{jaksch98}. Similar considerations apply 
for ultracold fermions: when there is in average no more than one particle 
per well and only the lowest band is occupied, a 
single-band tight-binding approximation works very well both when the 
dilute fermionic gas is polarized (corresponding to $a=0$) and when more 
species or levels of fermions are present. In the following we consider  
tight-binding models describing ultracold atoms in optical lattices, 
sticking to the non-interacting limit and discussing how to simulate 
artificial gauge potentials in such systems. We alert anyway the reader that, 
albeit non-rigorous, the experimental 
techniques to implement synthetic magnetic field 
are the same also for interacting particles, even though the interacting 
terms may modify or generate additional coefficients in the tight-binding model and 
introduce corrections to the results obtained with the Peierls substitution 
\cite{peierls33}.


To fix the notation, let us consider a lattice whose sites are 
denoted by $\vec{r}$: for a cubic $D$-dimensional 
lattice we have $\vec{r} \in \mathbb{Z}^D$. 
By using the Peierls substitution to take into account 
the effect of the magnetic field \cite{peierls33,luttinger51,kohn59}, 
the Hamiltonian we consider then reads
\begin{equation}
H(\{\phi\}) = -  \, 
\sum_{\vec{r} \, , \, \hat{j}}  w_{\hat{j}} \, c^{\dagger}_{\vec{r} + \hat{j}} 
\, e^{i \phi_{j}\left(\vec{r}\right)} 
 c_{\vec{r} } +  \mathrm{H.c.} \, ,
\label{peierls}
\end{equation}
where $\hat{j}$ are unitary vectors characterizing the links of the lattice, 
$w_{\hat{j}}$ are the hopping amplitudes 
(assumed isotropic in the following of the Section: $w_{\hat{j}} \equiv w$). 
The phases $\theta_{j}$ depend in general on the position $\vec{r}$ 
and can be thought as the integral of an artificial 
and classical vector potential $\vec{A}(\vec{r})$ between neighboring sites:
\begin{equation}
\phi_j(\vec{r})= \int_{\vec{r}}^{\vec{r}+\hat{j}} \vec{A}(\vec{x})\cdot d\vec{x} \, .
\label{peierls2}
\end{equation} 
The ladder operators $c_{\vec{r}}$ and $c^\dag_{\vec{r}}$ annihilate and create 
an atom in the lattice site $\vec{r}$ and they may obey 
either fermionic or bosonic commutation relations depending on the atoms 
species. 

It is important to emphasize that the artificial vector potential $\vec{A}$ 
constitutes a classical and static field; despite that, we can define $U(1)$ 
gauge transformations acting on the ladder operators of the previous 
Hamiltonian and on the vector potentials, which leave the dynamics of the system 
invariant; $\vec{A}$ is thus, in this context, a properly defined gauge 
potential (see the reviews \cite{dalibard11,goldman13rev} for more details). 
Notice as well that one can simulate gauge potentials without a periodic 
potential \cite{dalibard11,goldman13rev}, but that typically 
the implementation of magnetic potential in an optical lattice can 
crucially take advantage of the presence of the lattice potential itself 
(in other words, 
decreasing to zero the intensity of the laser beams amounts to make vanishing 
the magnetic potentials as well).

The effect of the gauge symmetry is that the main observables 
we must consider are gauge-invariant observables - 
although, due to the artificial nature of these gauge potentials, 
also gauge-dependent quantity may be evaluated in the experimental setups, 
going beyond the previous effective Hamiltonian description. 
The main gauge-invariant quantity determining the dynamics of the system 
is the magnetic flux which characterizes each plaquette in the lattice. 
Such flux describes an Aharonov-Bohm phase acquired by an atom hopping around 
a lattice plaquette and can be defined as:
\begin{equation}
\Phi_p = \sum_{(\vec{r},\hat{j}) \in p} \phi_{j}(\vec{r}) = \oint_p  \vec{A}(\vec{x})\cdot d\vec{x}\,,
\end{equation}
where $(\vec{r},\hat{j})$ labels the links along the plaquette $p$ in order to consider a counterclockwise path. The lattice spacing is denoted by $a$, 
and if not differently stated is intended to be set to $1$.

On the experimental side there are two broad classes 
of techniques which have been adopted to engineer 
effective Hamiltonians of the form in Eq. (\ref{peierls}). 
The first corresponds to the ``lattice shaking'', which consists 
in a fast periodic modulation of the optical lattice trapping the atoms 
whose effect is to reproduce, at the level of the slow motion of the atoms, 
the required complex hopping amplitudes. 
The second is the ``laser-assisted tunneling'' of the atoms in 
optical lattices in which the atom motion is suppressed along one 
direction and restored through the introduction of additional Raman lasers 
able to imprint additional space-dependent phases to the tunneling of 
the particles. Both these techniques allow for the generation of 
artificial magnetic fluxes and are based on non-trivial time-dependent 
Hamiltonians which determine, at the level of the slow motion of the system, 
a dynamics which can be described by an effective Hamiltonian of the kind in Eq. 
(\ref{peierls}).
In the following we will summarize first the technique developed in 
\cite{goldman14} which provides a very useful tool for the analysis 
of these driven time-dependent systems, and then we will describe some 
of the main examples of systems obtained through lattice shaking or 
laser-assisted tunneling.

\subsection{An effective description for periodically driven systems}

The technique for the analysis of periodically driven systems proposed 
in \cite{goldman14}, whose presentation we follow in this Section, 
is based on the distinction of two main ingredients whose 
combination describes the dynamics of modulated setups. 
The first is an effective Hamiltonian $H$, 
independent on the initial conditions of the dynamics, and capturing 
the long-time motion of the particles in the system. The second is a 
so-called kick-operator $K$ describing the effects due to 
the initial and final phases of the modulation. 
In particular it is responsible for both the initial 
conditions of the system and for the so-called micro-motion, 
which includes the periodic dynamics of all the fast-evolving degrees 
of freedom. To be explicit, let us assume that 
the modulated system is described by a time-periodic Hamiltonian 
$\tilde{H}(t)=\tilde{H}(t+T)$. It is then 
possible to decompose the evolution of the system into:
\begin{equation}
 U(t_i \to t_f) = e^{-iK(t_f)}e^{-iH(t_f-t_i)}e^{iK(t_i)}\,.
\end{equation}
Here $H$ is the effective, time-independent, Hamiltonian, 
not depending on $t_i$ and $t_f$; 
the kick operator $K(t)=K(t+T)$ is a periodic time-dependent operator, and hereafter we set $\hbar=1$. 
The approach in \cite{goldman14} consists in a series expansion 
of $H$ and $K$ in the small parameter $1/\omega =T/(2\pi)$, 
where $\omega$ is the driving frequency of the system and 
must constitute the largest energy scale of the problem.

To study optical lattices with non-trivial artificial magnetic fluxes 
it is usually enough to consider the long-term dynamics of the driven 
system and thus the effective Hamiltonian only 
(the situation would be different for systems involving also spin degrees 
of freedom, or for the evaluation of the heating of the driven system). 
To this purpose, we decompose the time-dependent Hamiltonian $\tilde{H}(t)$ 
into its Fourier component:
\begin{equation}
 \tilde{H}(t) = H_0 + \sum_{n>0} e^{i n\omega t} V^{(n)}+ \sum_{n>0} e^{-i n\omega t} V^{(-n)}
\end{equation}
with $V^{(-n)}=V^{(n)\dag}$. $H_0$ is the time-independent component of 
$\tilde{H}$, whereas the operators $V^{(-n)}$ are associated to its harmonics. 
In terms of these operators it is possible to show that 
the effective Hamiltonian reads:
\begin{eqnarray}
 H= H_0 &+& \frac{1}{\omega}\sum_{n=1}^\infty \frac{1}{n}\left[V^{(n)},V^{(-n)} \right] + \nonumber \\
 &+&\frac{1}{2\omega^2}\sum_{n=1}^\infty \frac{1}{n^2}\left( \left[ \left[V^{(n)},H_0 \right],V^{(-n)}\right] + {\rm H.c.}\right) + O(T^3) \,.  \label{hamgold}
\end{eqnarray}
This expansion allows for a determination of the effective Hamiltonian 
in the main examples of systems of ultracold atoms trapped 
in optical lattices, subject either to a modulation of the trapping lattice 
or two additional Raman couplings. One needs to consider carefully, 
though, the issue of the convergence of this series which must be evaluated 
specifically for each system. The readers are referred to \cite{goldman14} 
for more detail.

\subsection{Artificial gauge potentials from lattice shaking}

The first attempts to experimentally 
modify the hopping amplitudes of the effective 
tight-binding models for atoms trapped in optical lattices 
through the introduction of modulations date back to the works \cite{arimondo07,oberthaler08,arimondo09}. To understand how the effect of the lattice 
shaking can determine the tunneling amplitudes of the atoms let us first 
address a one-dimensional setup. Although, in this case, it is not possible to define magnetic fields and fluxes, the analysis of this simplified system will be useful to understand the appearance of artificial magnetic fluxes in higher dimensions.
We consider an optical potential of 
the form $V(t)=V_0\sin^2\left(x-\xi_0\cos(\omega t)\right)$ 
where $\omega$ is the shaking frequency. The Hamiltonian with this 
oscillating potential can be mapped in a co-moving frame 
$(x \to x + \xi_0\cos(\omega t))$ characterized by the following Hamiltonian:
\begin{equation}
 \tilde{H}(t)=p^2 + V_0\sin^2\left(x\right) -\frac{1}{2}x\xi_0\omega^2\cos(\omega t) \,,
\end{equation}
where the last term accounts for the additional force in the non-inertial 
frame and we set the mass to $m=1/2$ for the sake of simplicity. Finally, 
by approximating with a tight-binding model we obtain:
\begin{equation}
 \tilde{H}(t) =  \sum_x\left[-w  \, c^\dag_{x+1}c_x - w \, c^\dag_x c_{x+1} -\frac{1}{2}x\xi_0\omega^2\cos(\omega t)c^\dag_x c_x\right].
\end{equation}
In this case it is possible to derive the full effective Hamiltonian 
without recurring to a series expansion to obtain \cite{eckardt05}:  
\begin{equation} \label{hambessel}
 H=-w \mathcal{J}_0(\xi_0\omega/2) \sum_x c^\dag_{x+1}c_x + c^\dag_x c_{x+1}\,
\end{equation}
where $\mathcal{J}_0$ is a Bessel function of the first kind which 
renormalizes the tunneling amplitude and may assume either positive 
or negative values depending on $\xi_0\omega$. Despite the fact that, 
in this case, the hopping amplitude remains always real, 
it is interesting to notice that it can change sign.

In $1D$ systems, such a change of sign is simply translated in a different dispersion for the single particle problem; however, it is possible to extend this naive example to higher dimensions and less trivial geometries: in this case the result of the lattice shaking 
provides a first tool for the engineering of non-trivial fluxes. 
We also observe that, applying the formalism of \cite{goldman14}, we have 
that $H_0=-w \sum_x c^\dag_{x+1}c_x + {\rm H.c.}$, 
$V^{(1)}=V^{(-1)}=-x\omega^2 \, c^\dag_x c_x/4$ and all 
the other harmonics are absent. The effective Hamiltonian based on 
Eq. (\ref{hamgold}) would then correspond to a series expansion of 
the Bessel function in Eq. (\ref{hambessel}) \cite{goldman14}.

The lattice shaking techniques can be also extended to obtain complex hopping amplitudes. To this purpose, in this simple one-dimensional model, 
it is necessary to change the time-dependence of the modulation of the 
lattice in order to 
break the time-reversal symmetry [$V(t-t_0)=V(-t-t_0)$] and a shift 
antisymmetry [$V(t)=V(t+T/2)$] \cite{struck12}. This has been realized 
for the first time for a $Rb$ Bose-Einstein condensate 
and the presence of non-trivial hopping phases has been verified through 
time-of-flight measurements of its momentum distribution as a function of the 
modulation amplitude \cite{struck12}. 
This may be counterintuitive because, in one dimension, the observation of the hopping phase corresponds to the observation of a gauge-dependent quantity. We notice, however, that only the effective tight-binding models adopted in the description of the slow dynamics of the system, and the related observables, are indeed gauge-invariant; in the experiment, though, one can access also additional ``gauge-dependent'' observables through operations which do not have a physical 
counterpart in the effective toy model: this is the case of the time-of-flight imaging which follows from switching off the optical lattice. Such procedure maps the crystal momentum of the tight binding model (\ref{hambessel}) (which is a gauge-dependent quantity) into the velocity of the particles during the time of flight, which is an observable quantity which exists only considering the embedding of the system in the larger laboratory setting.

The effects of lattice shaking become even more remarkable in 
two-dimensional setups. In this case lattices may be accelerated along 
two different directions to cause a global periodic motion of the optical 
lattice around a closed orbit which yields, at the level of the effective 
Hamiltonian, a tight-binding model with non-trivial fluxes.

On triangular optical lattices this technique has been adopted to 
realize staggered flux configurations \cite{struck13} and, more recently, 
the same method, with a circular modulation of the lattice position 
\cite{oka09}, has been used to simulate the topological Haldane model 
\cite{haldane88} on the honeycomb lattice with a gas of fermionic 
$^{40}K$ \cite{esslinger14}. The Haldane model 
represents a topological insulator of fermions hopping in a 
honeycomb lattice with nearest-neighbor and next-nearest-neighbor tunnelings. 
It is based on both the presence of a pattern of staggered fluxes $\Phi$ 
to break time-reversal symmetry and an onsite staggering potential 
to break space inversion symmetry \cite{haldane88}. By varying the value of 
either of these parameters, the model undergoes topological phase transitions, 
characterized by discontinuities of the Chern number of the lowest 
energy band. These discontinuities have been experimentally detected 
through measurements of the drift of the center of mass of the system 
in the presence of an additional magnetic gradient to add an additional 
constant force \cite{esslinger14}. 

\subsection{Artificial gauge potential from laser-assisted tunneling}

Complex hopping amplitudes in the effective Hamiltonian can be obtained 
also through a different technique based on the introduction of pairs of 
Raman lasers coupling the low-energy states of the atoms trapped in the 
optical lattice. In this case the phase differences and space dependence 
of the Raman lasers may be inherited by the dynamics of the atoms, 
thus allowing for complex space-dependent amplitudes. 

To reach this result, however, it is necessary to first suppress 
the motion of the atoms in the optical lattice, at least, along one 
direction. This is obtained through the introduction of suitable energy 
offsets, depending on the positions, which shift the energy of neighboring 
sites by an energy $\Delta$. These offsets can be obtained, for example, 
by tilting the lattice (i.e., with gravity), or 
by introducing suitable magnetic gradients in the system 
which couple with the atomic magnetic dipole moments 
(thus generating a position dependent Zeeman term) or through the introduction of 
superlattices.

\begin{figure}
\includegraphics[width=7cm]{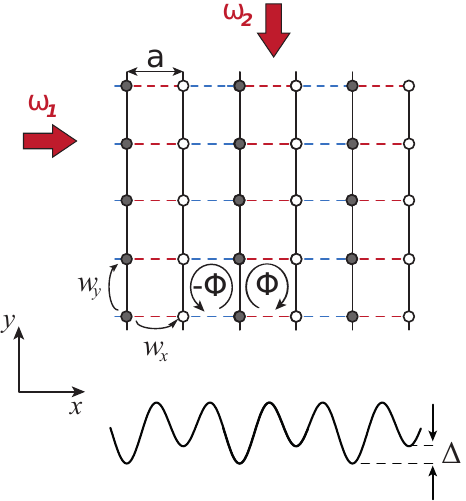} 
\caption{Schematic illustration of the setup adopted in the experiment 
\cite{aidelsburger11} for the realization of staggered magnetic fluxes. 
The tunneling along the horizontal direction is suppressed by the 
introduction of the staggering $\Delta$ through a superlattice. 
A pair of Raman lasers (red arrows) are added to the system to restore 
the horizontal tunneling. As a result the even horizontal links 
(blue dashed links) acquire a tunneling phase $e^{-i\vec{k}_R \vec{r}}$, 
whereas the odd (red dashed links) acquire the opposite phase 
$e^{i\vec{k}_R \vec{r}}$. $\vec{k}_R$ is the recoil momentum of the pair of 
Raman lasers.} \label{fig:setup}
\end{figure}

Let us start by considering one of the simplest realization of strong fluxes 
in optical square lattices as experimentally realized \cite{aidelsburger11}. 
In this experiment an optical superlattice, generated by a standing 
wave with wavelength $2a$, was used to introduce an additional staggering 
along the $\hat{x}$ direction for the trapped atoms, such that the initial, 
time-independent setup can be modeled by the following Hamiltonian:
\begin{equation}
 \tilde{H}'_0= -  w \, 
\sum_{\vec{r} \, , \, \hat{j}}  \left(  c^{\dagger}_{\vec{r} + \hat{j}} c_{\vec{r} } +  \mathrm{H.c.}\right) + \frac{\Delta}{2}\sum_{\vec{r}} (-1)^x c^\dagger_{\vec{r}}c_{\vec{r} } \,
\end{equation}
where $\vec{r}=(x,y)$ and $\Delta$ is the staggering related 
to the amplitude of the superlattice (see Fig. \ref{fig:setup}). 
Two running Raman lasers with wave vectors $\vec{k}_{1,2}$ and 
frequencies $\omega_{1,2}$, tuned such that $\omega_1-\omega_2=\Delta$, 
are then introduced in the system. The associated electric field is 
$E_1 \cos(\vec{k_1}\vec{r}-\omega_1 t) + E_2\cos(\vec{k_2}\vec{r}-\omega_2 t)$ 
which, neglecting the fast oscillating terms, generates a potential:
\begin{equation}
 V(t) = \kappa e^{i\left(\vec{k}_R \vec{r} -\Delta t \right) }c^\dagger_{\vec{r}}c_{\vec{r}} + {\rm H.c.}\,,
\end{equation}
where $\kappa=2E_1E_2$ and 
$\vec{k}_R= \vec{k}_1 - \vec{k}_2$ is the recoil momentum of the Raman lasers.

The time-evolution of the system is ruled by the Schr\"odinger equation 
$i  \partial_t \psi = \tilde{H}'(t) \psi$, 
where $ \tilde{H}'(t)=\tilde{H}'_0+ V(t)$. 
Since the static Hamiltonian $\tilde{H}'_0$ contains the staggered-potential 
term that explicitly diverges with the driving frequency $\Delta$, 
it is convenient to apply the unitary transformation \cite{goldman14b,mazza15}
\begin{equation}
\psi =  R(t) \tilde \psi = \exp \left (\! -i   W t \right ) \tilde \psi, \label{transf_hof}
\end{equation}
with $W$ being the staggering term:
\begin{equation}
 W = \frac{\Delta}{2}\sum_{\vec{r}} (-1)^x c^\dagger_{\vec{r}}c_{\vec{r} }    \,.
\end{equation}
Such transformation removes the diverging term and maps $\tilde{H}'$ into:
\begin{eqnarray}
 \tilde{H}(t)=R^{\dagger}(t) \left[ \tilde{H}'_0 +  V(t) \right]  R(t) -  W= \\
 =H_0 + V^{(1)} e^{i\Delta t} + V^{(-1)}  e^{-i\Delta t},
\end{eqnarray}
where
\begin{eqnarray}
H_0 &=& -w \sum_{x,y}\left(  c^\dag_{x,y+1} c_{x,y} +c^\dag_{x,y}c_{x,y+1}\right) ,\\
 \hat V^{(1)} &=& \kappa \sum_{\vec{r}} e^{-i\vec{k}_R \vec{r}}c^\dagger_{\vec{r}}c_{\vec{r}} - w \sum_{x \, {\rm odd},y} \left(c^\dag_{x+1,y} c_{x,y} +  c^\dag_{x-1,y} c_{x,y} \right) , \\
 \hat V^{(-1)} &=& \kappa \sum_{\vec{r}} e^{i\vec{k}_R \vec{r}}c^\dagger_{\vec{r}}c_{\vec{r}} - w \sum_{x \, {\rm even},y} \left(c^\dag_{x+1,y}c_{x,y} + c^\dag_{x-1,y}c_{x,y} \right) .\label{theVeq}
\end{eqnarray}
From these terms it is easy to derive the effective Hamiltonian in 
Eq. (\ref{hamgold}) at first order:
\begin{eqnarray}
 &H&=-w \sum_{x,y}\left(  c^\dag_{x,y+1} c_{x,y} +c^\dag_{x,y}c_{x,y+1}\right) -\\ \nonumber
  &-&\frac{w \, \kappa}{\Delta}\sum_{x \, {\rm even},y} \left[ \left(e^{-i\vec{k}_R \hat{x}} -1 \right) \left( e^{-i\vec{k}_R \vec{r}}c^\dag_{x+1,y}c_{x,y} +  e^{i\vec{k}_R \vec{r}}c^\dag_{x,y}c_{x-1,y}\right)  +{\rm H.c.}\right]  + O(1/\Delta^2)\,,
\end{eqnarray}
This effective Hamiltonian describes in general a 
two-dimensional model with staggered magnetic fluxes where the sign of 
the fluxes alternate in the plaquettes belonging to even and odd columns. 
In the experiment \cite{aidelsburger11}, the recoil momentum was 
chosen as $\vec{k}_R =(\hat{x}+\hat{y}) \Phi $. In this case the 
$\hat{x}$ component has no relevance in the definition of the 
fluxes and the previous Hamiltonian becomes, after a suitable gauge 
transformation:
\begin{eqnarray}\label{ham_1}
 &H&=-w_y \sum_{x,y}\left(  c^\dag_{x,y+1} c_{x,y} +c^\dag_{x,y}c_{x,y+1}\right)  -\\ \nonumber
 &\phantom{H}&w_x \sum_{x \, {\rm even},y}\left[ \left( e^{-i\Phi y}c^\dag_{x+1,y}c_{x,y} +  e^{i \Phi y}c^\dag_{x,y}c_{x-1,y}\right)  +{\rm H.c.}\right]+ O(1/\Delta^2)
\end{eqnarray}
with $w_y=w$ and $w_x=2 w \kappa\sin(\Phi/2)/{\Delta}$ 
(this value is the one obtained at the first order in the perturbative 
expansion, and it must be considered only an approximation). From 
the Hamiltonian in Eq. (\ref{ham_1}) it is evident 
the alternation of fluxes $\pm \Phi$ on the plaquettes along the 
horizontal direction.

The introduction of the staggering term, however, allows 
also for more refined setups in which the even and odd links may be separately 
addressed \cite{goldman14b}. This requires the introduction of two 
different pairs of Raman lasers with opposite frequency shifts 
$\pm \Delta$ and it permits to obtain systems 
with a uniform magnetic flux $\Phi$ in each plaquette \cite{goldman14b}. 
In this way the Hofstadter model on the square lattice has been realized 
for $^{87}Rb$ \cite{bloch15} and it was possible to measure the Chern number 
of the different energy bands through the motion of the mass center 
of the system.

The staggered potential, however, it is not the 
only possible choice to suppress the motion along one direction. 
The first quantum simulations with ultracold atoms of the Hofstadter model 
\cite{Aidelsburger2013,Miyake2013} were instead based on an external 
potential of the kind $W=\sum_{\vec{r}}\Delta x c^\dag_{\vec{r}}c_{\vec{r}}$. 
In this case the introduction of two Raman lasers yields indeed to 
an effective Hamiltonian with rectified fluxes, and this result 
can be obtained with calculations analogous to the previous one 
where the distinction between even and odd links is no longer required, 
and all the horizontal links acquire a tunneling phase of 
the form $e^{i\vec{k}_R \vec{r}}$ consistent with a constant flux.

The laser-assisted techniques to design artificial gauge potentials 
are extremely versatile, and the previous approach can be generalized 
to different geometries and to multi-component species. The introduction 
of additional potential through superlattices, for example, enabled 
the realization of ladder models pierced by uniform fluxes 
which are characterized by chiral currents and a Meissner-like 
effect \cite{bloch14}. Furthermore the introduction of 
spin-dependent potentials, as in \cite{Aidelsburger2013}, 
permits to mix different spin-species subject to opposite magnetic 
fluxes and some theoretical proposals generalized these systems to 
engineer an artificial spin-orbit couplings for two-component atoms 
\cite{goldman14,mazza15}.

\section{Cubic lattice tight-binding models with commensurate flux}
\label{Abelian}

In this Section we consider cubic lattices in a magnetic field, focusing 
on the case of Abelian fluxes. This is  
a very interesting mathematical problem in itself and it has a counterpart 
in the experimental implementations we discussed in the previous Section.

\subsection{Isotropic flux}
\label{isot}

We start our study with  a single-species tight-binding model 
on a cubic lattice with $N= L^3$ sites, in the presence of an Abelian 
uniform and static magnetic field $\vec{B} =  \Phi \, (1,1,1)$ isotropic 
on the three directions. This field gives rise on each plaquette 
of the lattice (with area $a^2$, $a$ being the lattice spacing) 
to a magnetic flux $\Phi = \vec{B} \cdot \vec{a}^2$. The presence of this 
flux amounts to a phase $e^{i \Phi}$ gathered by a particle hopping around 
a single plaquette, because of the Stokes theorem. 
The generalization of this model to many species is straightforward, 
since an Abelian gauge field does not mix the different species. 
The magnetic flux is chosen commensurate: 
\begin{equation}
\Phi = 2 \pi \frac{m}{n} \, .
\end{equation} 
The commensurate condition allows us to solve the model analytically, 
imposing periodic boundary condition on the lattice. At variance, 
in the incommensurate case the determination of spectrum requires 
a numerical solution of the real space tight-binding matrix, see e.g. 
\cite{lin96,bruning04} - for a discussion of the Hofstadter butterfly 
in three dimensions see \cite{koshino01}, while a study in higher dimensions 
is reported in \cite{kimura14}.

By exploiting the gauge redundancy, the static magnetic field $\vec{B}$ 
can be associated to various physically equivalent gauge potentials 
$A_{\mu} (\vec{x})$. For the sake of simplicity, we choose here a 
time-independent gauge configuration, adopting the (static) Coulomb gauge 
$A_0(\vec{x}) =0$.

Because of the magnetic phases in Eq. (\ref{peierls2}), 
the sites of the lattice, all equivalent each other at $\vec{B} = 0$, 
get inequivalent, the inequivalence lying in the phases gathered after 
each hopping along the bonds starting from a certain site. In this way, 
the lattice gets divided, in a gauge-dependent way, 
in a certain number of sublattices. Exploiting this freedom,
in order to perform calculations in the easiest way as possible, it is useful 
to look for the (set of) gauge(s) characterized, for a given 
commensurate magnetic flux $\Phi = 2 \pi \frac{m}{n}$, by 
the smallest number of sublattices.

A simple (and still not unique) gauge fulfilling this requirement is 
\cite{Hasegawa}:
\begin{equation} 
\vec{A}= \frac{2 \pi}{a^2} \frac{m}{n}   
 \, (0, x-y, y-x) \, ,
\label{potgenab}
\end{equation}
with permutations in ${x,y,z}$ also equally acceptable. 
This gauge is a three-dimensional generalization of the 
Landau gauge in two dimensions \cite{LandauQM}, reducing indeed 
to the Landau gauge in this limit ($w_{\hat{z}} \to 0$), 
up to a gauge redefinition to absorb the term $-y$ in $A_y$. 

Assuming the choice in Eq. (\ref{potgenab}),
the Hamiltonian in Eq. (\ref{peierls}) can be recast in the form  
\begin{equation} 
 H =  - \sum_{\vec{r}} \left[ w_{\hat{x}} \, c^\dag_{\vec{r}+\hat{x}} \, c_{\vec{r}} + w_{\hat{y}} \, U_{\hat{y}} (x,y) \, 
c^\dag_{\vec{r} + \hat{y}}c_{\vec{r}} +  w_{\hat{z}} \, U_{\hat{z}} (x,y) \, 
c^\dag_{\vec{r} + \hat{z}}c_{\vec{r}}  \right] + \, \mathrm{h.c.} \, ,
\label{Hab}
\end{equation}
where the tunneling magnetic phases $U_{\hat{j}} (x,y) =e^{i \, \phi_{\hat{j}}(x,y)}$ are defined as:
\begin{eqnarray}
& U_x = 1 \, , \label{ux}\\
 &U_{\hat{y}} (x,y) = \exp\left( {i \, \int_{x,y,z}^{x,y+a, z}A_y \, \mathrm{d}y}\right) =\exp\left( {i \, 2 \pi \, \Big( \frac{x - y}{a} - \frac{1}{2} \Big) \, \frac{m}{n} } \right) \, , \label{uy} \\
 &U_{\hat{z}} (x,y) =  \exp\left( {i \, \int_{x,y,z}^{x,y, z+a}A_z \, \mathrm{d}z}\right) = \exp\left( { - i \, 2\pi \, \Big( \frac{x - y}{a} \Big) \,  \frac{m}{n} } \right)  \, . \label{uz}
\end{eqnarray}
The hopping phases in Eqs. (\ref{uy}) and (\ref{uz}) explicitly depend 
on the positions labelled modulo $n$. Since 
the $z$ coordinate is not present in Eq. (\ref{potgenab}), 
every wave-function of the Hamiltonian in Eq. (\ref{Hab}) can be written as 
\cite{LandauStat2}:
\begin{equation}
 \psi(x,y,z)=e^{ik_z z}  \, u (x,y) \, ,
\end{equation}
allowing for a dimensional reduction of the eigenvalues problem as for the 
corresponding problem in two dimensions where the Harper equation is found 
\cite{hofstadter76,thouless90}.

The Hamiltonian in Eq. (\ref{Hab}) with the hopping phases in Eqs. 
(\ref{ux})-(\ref{uz}) is translational invariant. For gauge invariant systems, 
translational invariance implies that a translation of the coordinates by a 
vector $\vec{w}$ transforms the Hamiltonian of the system 
to a gauge-equivalent one \cite{LandauStat2}, and one may find 
\cite{lepori2015}:
\begin{equation} 
H(\vec{r} + \vec{w})= \mathcal{T}_{\vec{w}}^\dag(\vec{r}) H(\vec{r}) \mathcal{T}_{\vec{w}} (\vec{r})   \, ,
\label{invab}
\end{equation}
with $\mathcal{T}_{\vec{w}} (\vec{r})\in U(1)$ being a 
suitably chosen local gauge transformation which depends on $\vec{w}$.

The Hamiltonian in Eq. (\ref{Hab}) with the gauge potential in 
Eq. (\ref{potgenab}) is translationally invariant because it 
fulfills Eq. (\ref{invab}). 
We stress that the potentials of the form in Eq. (\ref{potgenab}) are
 not the only ones satisfying the condition in Eq. (\ref{invab}), but 
instead all the potentials obtained from Eq. (\ref{potgenab}) 
through local gauge transformation are characterized by the same 
physical translational invariance. The property in Eq. (\ref{invab}) 
is indeed a physical property of the system which is reflected on 
all the gauge-invariant observables, as for example, the Wilson loops
$W(\mathcal{C}) = {\bf P}e^{i\oint_{\mathcal{C}} \vec{A}(\vec{r})\cdot d\vec{r}}$ 
evaluated on closed paths along the lattice. 

The Hamiltonian in Eq. (\ref{Hab}) is also periodic with period $n$ along 
$\hat{x}$ $\hat{y}$, due to the presence of the nontrivial magnetic 
hopping phases $U_{\hat{y}} (x,y) \, , U_{\hat{z}} (x,y)$, 
thus the reduced wavefunctions $u (x,y)$ have the same periodicity. 
In this way the magnetic unitary cell, defined by the elementary translations 
leading from a site to equivalent ones in the three lattice directions, 
can be defined now as enlarged $n$ times along two directions 
(say $\hat{x},\hat{y}$). 

The problem to find the eigenvalues of the Hamiltonian in Eq. (\ref{Hab}) 
on a lattice with $N$ number of sites would na\"ively require in general the 
diagonalization of a $N \times N$ adjacency matrix in real space. 
However, assuming translational invariance, the calculation can be 
remarkably simplified by exploiting the division in sublattices seen above.
Indeed, in the presence of a magnetic flux $\Phi = 2\pi\frac{m}{n}$ 
and working in the gauge in Eq. (\ref{potgenab}), 
the cubic lattice divides in $n$ sublattices, labelled by 
the quantity $(x-y) \, \mathrm{mod} (n)$. 
In this way the Hamiltonian in Eq. (\ref{Hab}) becomes:
\begin{equation}
H = -  \sum_{\hat{j}} w_{\hat{j}} \sum_{s} e^{i \phi_{s, \hat{j}}}  \sum_{\vec{r}_{s}}  \, c^{\dagger}_{\vec{r}_{s} + \hat{j}} \, c_{\vec{r}_{s}} \, + \mathrm{H.c.} \, ,
\label{peierls3}
\end{equation}
where $s$ labels the sublattices and $\vec{r}_s$ labels the sites of the 
$s$-th sublattice. 
 
Since the magnetic unitary cell is defined now as enlarged $n$ times 
along two directions, the corresponding magnetic Brillouin zone (MBZ) 
in momentum space becomes 
$\{ \frac{2 \pi }{a}, \frac{2 \pi }{a n}, \frac{2 \pi }{n a} \}$ 
[or other permutations of the factors $\frac{1}{n}$ between 
the space directions, consisting in a mere redefinition of 
$\vec{A}(\vec{r})$]. Correspondingly, the allowed momenta 
$\vec{k}$ are $\frac{N}{n^2}$, of the form 
$\vec{k} = \{ \frac{2 \pi }{a L_x } \, p_x, \frac{2 \pi }{a L_y} \, 
p_y, \frac{2 \pi }{a L_z} \, p_z \}$, with $p_{\hat{x}} = 0, \dots, L-1$ and 
$p_{\hat{y},\hat{z}} = 0, \dots, \frac{L}{n}-1$.
In this way, all the physical quantities display a 
momentum periodicity $\vec{k} \to \vec{k} + \frac{2 \pi}{n a} 
(l_x, l_y,  l_z)$, with $l_i$ being again integer numbers. 

The counting of the $\frac{N}{n^2}$ allowed momenta proceeds as follows: 
each sublattice differs from another one by $\pm \hat{x}$ or $\pm \hat{y}$ 
translations that are not primary, then we 
correspondingly expect $n$ sets of $\frac{N}{n}$ 
inequivalent energy eigenstates \cite{LandauStat2}. 
These sets form $n$ subbands in the MBZ. However, any sublattice 
is further divided in $n$ sub-sublattices differing by a primary 
translation $\pm (\hat{x} + \hat{y})$, leaving invariant the potential 
in Eq. (\ref{potgenab}). In this way, 
any $\frac{N}{n}$-fold set of eigenstates is again partitioned in 
$n$ equivalent and degenerate sub-sets, each one having 
$\frac{N}{n^2}$ element. These elements are parametrized by the 
$\frac{N}{n^2}$ momenta in the MBZ described above. Moreover the 
second partition translates in $n$-fold degeneracy of each subband.
In the particular case $\Phi = \pi$, two sub-bands are obtained, 
touching in Weyl cones as discussed in \cite{lepori2015,LMT,ketterle2015}; 
in this case the system in Eq. (\ref{Hab}) is the direct 
three-dimensional generalization of the square lattice model with 
$\pi$-fluxes in \cite{affleck}.

The Hamiltonian in Eq. (\ref{peierls3}) can be expressed 
in momentum space using 
the formulas $c_{\vec{r}_s} = \frac{1}{\sqrt{N/n^2}} \, \sum_{\vec{k}} 
\, c(\vec{k}) \, e^{i \vec{k} \cdot \vec{r}_{s}} $ and 
$\sum_{\vec{r}_{s}}^{\prime} \, e^{i \vec{k} \cdot \vec{r}_{s}} = \frac{N}{n^2}$. 
The vectorial label $\vec{r}_s$ runs here on the $\frac{N}{n^2}$ sites 
of one sub-sublattice of the sublattice $s$, the upper index in the sum 
of the second formula meaning this restriction. We obtain:
\begin{equation}
H = -  \sum_{\vec{k}} \sum_{\hat{j}} w_{\hat{j}} \sum_{s} e^{i \phi_{s, \hat{j}}} \,  e^{-i \vec{k} \cdot \vec{j}} \, c^{\dagger}_{s^{\prime} = s + \hat{j}} (\vec{k}) \, c_{s} (\vec{k})\, + \mathrm{H.c.} \, ,
\label{peierls4}
\end{equation}
where we have also taken into account that, starting from the 
$s$-th sublattices and moving in the $\hat{j}$ direction, a new sublattice 
(denoted as $s^{\prime}$) is univocally found, 
as consequences of Eq. (\ref{potgenab}). 
This is the reason oh the notation $s^{\prime} = s + \hat{j}$.

The Hamiltonian in Eq. (\ref{peierls4}) 
can be recast in the sublattices basis as:
\begin{equation}
H = -  \,  \sum_{\vec{k}} \sum_{\hat{j}} w_{\hat{j}} \sum_{s}  \, 
c^{\dagger}_{s^{\prime} = s + \hat{j}}(\vec{k})  \Big(T_{\hat{j}}^{\mathrm{AB}} \Big)_{s^{\prime}, s} \,  e^{-i \vec{k} \cdot \vec{j}} \, c_{s}(\vec{k}) \, + \, \mathrm{H. c.} \, ,
\label{hamab}
\end{equation}
where 
$T_{\hat{j}}^{\mathrm{AB}}$ are $n \times n$ hopping matrices in the sublattice basis, reading, up to ciclic permutations ${\bf Z}_n$:
\begin{equation}
\begin{array}{ccccc}
T_{\hat{x}}^{\mathrm{AB}} = \left(
      \begin{array}{cccc}
        0 & 1 & \ldots & 0\\
        0 & 0 & \ddots & 0\\
        0 & 0 & \ldots & 1\\
        1 & 0 &  0 & 0\\
             \end{array} \right)

      T_{\hat{y}}^{\mathrm{AB}} = e^{- i \pi \frac{m}{n}} \, \left(
      \begin{array}{cccc}
        0 & \ldots & 0 & \varphi_0\\
       \varphi_{1}  & 0 & \ldots & 0\\
        0 & \ddots & 0 & 0\\
         0 & \ldots &  \varphi_{n-1} & 0\\
      \end{array} \right)   \,

        T_{\hat{z}}^{\mathrm{AB}} =
                \left(
      \begin{array}{cccc}
        \varphi_0 & 0 & \ldots & 0\\
        0 & \varphi_{n-1} & 0 & 0\\
        0 & 0 & \ddots & 0\\
        0 & 0 &  \ldots & \varphi_{1}\\
      \end{array} \right) 
\end{array} \, ,
\label{matab}
\end{equation}
with $\varphi_{l} = e^{i 2 \pi \frac{m}{n} l} \, , \, l = 0, \dots , n-1 $.

The matrices $T_{\hat{x, y,z}}^{\mathrm{AB}}$ do not commute each other, 
implying that the unitary cell of the lattice does not coincides with 
its geometric smallest cell (with area $a^2$), as seen above. 
However, the $n$-power of these matrices yields the identity: 
$\left(T_{\{\hat{x},\hat{y},\hat{z}\}}^{\mathrm{AB}}\right)^n = {\bf 1}_{n \times n} $, 
recovering the MBZ $\{ \frac{2 \pi }{a}, \frac{2 \pi }{a n}, 
\frac{2 \pi }{a n} \}$. Moreover, if $\frac{m}{n} \neq \frac{1}{2}$ 
the unitary matrices $T_{\hat{x, y,z}}^{\mathrm{AB}}$  are not invariant 
(even possibly up to a global phase) by the conjugate operation, 
reflecting the breaking of the time-reversal symmetry,  due to the magnetic 
field $\vec{B}$ itself. 
A notable result of the discussion above is that the diagonalization 
of a $N \times N $ matrix is reduced to the diagonalization of a 
$n \times n $ one.

We conclude this Section by observing that in the presence of 
more species (labelled by the index $\alpha = 1, \dots , m$) 
hopping on the lattice and subject to the Abelian gauge potential 
in Eq. (\ref{potgenab}), the Hamiltonian in Eq. (\ref{hamab}) generalizes to
\begin{equation}
H = - \sum_{\vec{k}}  \sum_{\hat{j}} w_{\hat{j}} \sum_{s} \, 
c^{\dagger}_{s^{\prime} = s + \hat{j}, \alpha^{\prime}} (\vec{k}) \left( T_{\hat{j}}^{\mathrm{AB}} \otimes  {\bf 1}_{m \times m} \right)_{s^{\prime}, \alpha^{\prime}, s, \alpha} \,  e^{-i \vec{k} \cdot \vec{j}} \, c_{s , \alpha} (\vec{k}) \, + \, \mathrm{H. c.} \, ,
\label{hamab2}
\end{equation}
where no mixing of the different species involved occurs.

\subsection{Generalization: anisotropic Abelian lattice fluxes}

In the previous analysis we assumed that the value of the fluxes was the same for the three orientations of the plaquettes. Now we discuss some extension in which we relax this hypothesis. 
We analyze first the case in which the magnetic field $\hat{z}$ is 
perturbed, such that we introduce an anisotropy in the previous system:
\begin{equation}
\vec{B} = \frac{2 \, \pi}{a^2} \, \Big(\frac{m}{n}, \frac{m}{n},  \frac{m_z}{n_z} \Big)\, .
\label{B}
\end{equation}
Again we may assume, without any loss of generality, 
$m_z$ and $n_z$ prime with each other as well as $m$ and $n$. 
In the case of Eq. (\ref{B}), a gauge similar to the one in 
Eq. (\ref{potgenab}) can be used:
\begin{equation}
\vec{A}= \frac{2 \pi}{a^2}  
 \, \Big(0, \frac{m_z}{n_z} \, (x-y), \frac{m}{n}  \, (y-x) \Big) \, .
\label{potgenab2}
\end{equation}
This choice still depends on one parameter only, 
thus ensuring the appearance of minimal gauge-dependent sublattices.  
More in detail, the lattice divides again in 
$n_2 = \mathrm{l. c. m. }(n,n_z)$ inequivalent sublattices, 
defined by the periodicity of the phases in the 
$\hat{x}$ and $\hat{y}$ directions and labelled by the set 
$(x-y) \mathrm{mod} (n_2)$.
Indeed the hopping phases from Eq. (\ref{potgenab2}) are:
\begin{equation}
\phi_{\vec{r} + \hat{j} , \vec{r}} =  \frac{2 \pi}{a}  
 \, \Big(0, \frac{m_z}{n_z} \, \Big(x-y-\frac{a}{2}\Big), \frac{m}{n}  \, (y-x) \Big) \, .
\end{equation}
Similarly to the previous case, each sublattice is again 
divided in $n_2$ equivalent sub-sublattices, each one having
$\frac{N}{n_2^2}$ sites. For this reason, similarly to the Subsection 
\ref{isot}, $n_2$ subbands appear, each of them with a $n_2$ degeneracy. 
Correspondingly the BZ divides by $n_2$ in two directions, so that  
$\vec{k} = \{ \frac{2 \pi }{a L_x } \, p_x, \frac{2 \pi }{a L_y} \, p_y, 
\frac{2 \pi }{a L_z} \, p_z \}$, $p_{\hat{x}} = 0, \dots, L-1$, 
$p_{\hat{y},\hat{z}} = 0, \dots, \frac{L}{n_2}-1$, or permutations in 
the pairs of the restricted momentum directions. 
The Hamiltonian in Eq. (\ref{peierls}) can then be rewritten, 
in terms of these quasi-momenta, as in Eq. (\ref{hamab}), 
by means of  three $m_2 \times m_2$ matrices in the basis of the $m_2$ 
sublattices, derived similarly to the ones in Eq. (\ref{matab}).

In the completely asymmetric case the magnetic potential reads
\begin{equation}
\vec{B} = \frac{2 \, \pi}{a^2} \, \Big(\frac{m_x}{n_x}, \frac{m_y}{n_y},  \frac{m_z}{n_z} \Big) \, ,
\label{basim}
\end{equation}
a convenient gauge choice, inducing 
the magnetic field in Eq. (\ref{basim}), reads:
\begin{equation}
\vec{A}_{{\bf AB} }= \frac{2 \pi}{a^2}  
 \, \Bigg(\Big(\frac{m_y}{n_y}-\frac{m_x}{n_x} \Big) \, (z-x), \frac{m_z}{n_z} \, (x-y), \frac{m_x}{n_x} \, (y-x) \Bigg) \, .
\label{potgenab3}
\end{equation}
The hopping phases from Eq. (\ref{potgenab3}) are:
\begin{equation}
\phi_{\vec{r} + \hat{j} , \vec{r}} =  \frac{2 \pi}{a}  
 \, \Bigg(\Big(\frac{m_y}{n_y}-\frac{m_x}{n_x} \Big) \, \Big(z-x-\frac{a}{2}\Big), \frac{m_z}{n_z} \, \Big(x-y-\frac{a}{2}\Big), \frac{m_x}{n_x}  \, (y-x) \Bigg) \, .
\end{equation}
The gauge in Eq. (\ref{potgenab3}), similarly to the previous case, ensures 
the appearance of the minimum number of gauge-dependent sublattices. 
In particular, due to the simultaneous $x$ dependence of all the components 
of $\vec{A}_{\mathrm{AB}}$  and following the same logic as in the 
Subsection \ref{isot}, we obtain  
\begin{equation}
n_s = \mathrm{l. c. m. }(n_x, n_y, n_z)
\end{equation}
inequivalent sublattices 
(obtained varying $y$ and $z$ at fixed $x$) and corresponding subbands. 
Again each sublattice is then further divided in equivalent subsublattices. 

More in detail, the counting of these subsublattices proceeds as follows. 
Starting from a point $(x_0, y_0, z_0)$ belonging to a certain sublattice, 
they are obtained by adding $1$ to each components: 
$(x_0, y_0, z_0) \to (x_0 +1, y_0 +1, z_0+1)$. The variable $z$ has 
periodicity given by $\mathrm{l. c. m. }(n_x,n_y)$ possible
inequivalent values, $y$ has periodicity $\mathrm{l. c. m. }(n_x,n_z)$ 
and finally $x$ has $\mathrm{l. c. m. }(n_x,n_y,n_z)$ inequivalent values. 
For this reason each inequivalent sublattice divides in
\begin{eqnarray}
n_d = \mathrm{min} \Big(\mathrm{l. c. m. }(n_x,n_y) \, , \,  \mathrm{l. c. m. }(n_x,n_z) \, , \,  \mathrm{l. c. m. }(n_x,n_y,n_z) \Big) = \\ \nonumber
= \mathrm{min} \Big(\mathrm{l. c. m. }(n_x,n_y) \, , \,  \mathrm{l. c. m. }(n_x,n_z) \Big)
\end{eqnarray}
equivalent subsublattices. 

Correspondingly, we find  $\frac{N}{n_s \times n_d}$ 
quasi-momenta defining each subband: 
$\vec{k} = \{ \frac{2 \pi }{a L_x } \, p_x, \frac{2 \pi }{a L_y} \, p_y, \frac{2 \pi }{a L_z} \, p_z \}$, $p_{\hat{x}} = 0, \dots, L-1$, $p_{\hat{y}} = 0, \dots, \frac{L}{n_s}-1$, $p_{\hat{z}} = 0, \dots, \frac{L}{n_d}-1$, or 
permutations in the pairs of the restricted momentum directions. 
The Hamiltonian  in Eq. (\ref{peierls}) can be then rewritten, 
in terms of these quasi-momenta and in the basis of the $n_s$ sublattices 
as in Eq. (\ref{hamab}), by means of three $n_s \times n_s$ matrices similar 
to the ones in Eq. (\ref{matab}).

\section{Two applications of synthetic gauge 
potentials in $1D$ rings}
\label{long_phi}

The possibilities offered by ultracold atoms in optical lattices 
to engineer tight-binding models in tunable magnetic potential open as well 
new possibilities also in the field of quantum information in the sense 
that they could be used in perspective 
to perform quantum information tasks and control 
the amount of entanglement of the system. Here, as two examples we believe 
paradigmatic of such potentialities, 
we want to shortly address two specific applications 
showing how tuning a gauge potential could modify the capability of 
a system to share quantum information. We consider one-dimensional 
models of free fermions on a ring geometry 
in the presence of a synthetic magnetic field piercing 
the ring. 
We first analyze for short-range lattice models how 
a topological phase helps to enhance the fidelity in a quantum state transfer 
(QST) process between different sites of the lattice 
\cite{Paganelli2009,Apollaro2015}. Then we study a long-range model 
to see how the presence of a topological phase can lead to the a volume-law 
behavior of the entanglement entropy (EE) for the ground state of the system.

\subsection{Quantum state transfer in a ring pierced by a magnetic flux}

We consider a one-dimensional tight-binding model for 
free fermions with nearest-neighbors hopping 
in a ring geometry embedded in a magnetic 
field. Such magnetic field determines the boundary conditions of the problem: its role is to induce an Aharonov-Bohm phase in the transport of a particle along a full circle around the ring geometry.
The Hamiltonian of the system reads
\begin{equation}
H=-w \sum_j  e^{i \phi }  c^\dagger_j c_{j+1} + h.c. \, ,
\end{equation}
where $\phi=\frac{2\pi}{N_S} \Phi$  ($\Phi$ being the Abelian magnetic 
flux piercing the ring chain in units of $2 \, \pi$), 
$N_S$ is the number of sites, and the site coordinate is $r=a j \bmod N_S $. 

The single-particle energy dispersion is
\begin{equation}
E_k (\phi)=-2w  \cos\left(a k -\phi  \right)=2t \cos\left(\frac{2\pi }{ N_S } (n_k-\Phi) \right) \, ,
\end{equation}
with $k=\frac{2\pi }{a N_S } \, n_k$. 
Due to the nontrivial phase $e^{i \phi }$, the single-particle 
dispersion is shifted, as well as the points corresponding to the Fermi surface.
For a single fermion, the introduction of this topological phase 
affects the wave-packet diffusion, giving a useful tool to
optimize the quantum state transfer of a 
certain state from a part of the chain to another one.

\begin{figure}[htbp]
\begin{center}
\includegraphics[height=8cm,angle=-90]{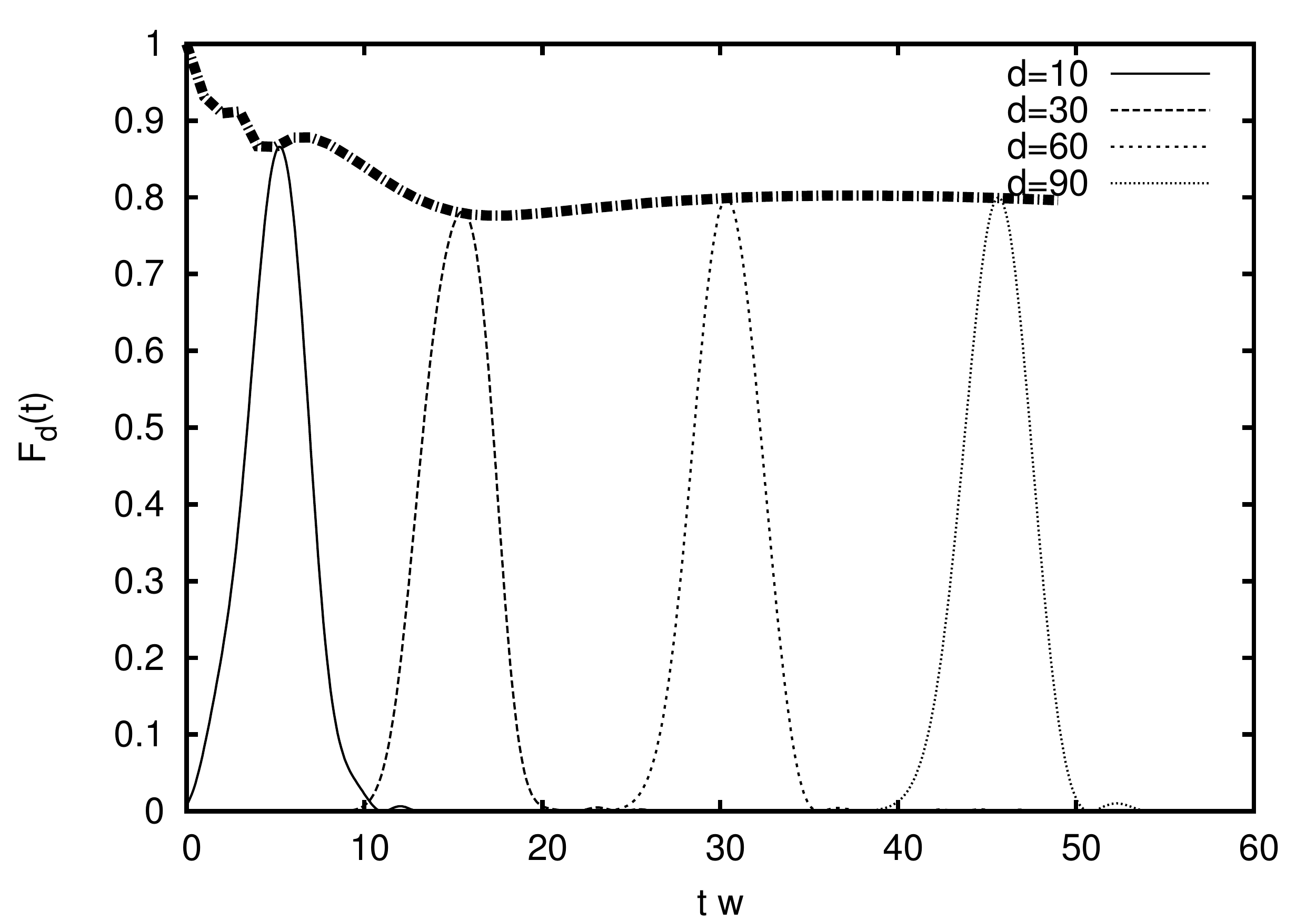}
\end{center}
\caption{Time evolution of the fidelity $F_d(t)$ in the square 
packet preparation ($M=5$), with phase $\phi=-\pi/2$, 
different values for the final site $d=10,30,60,90$ and $N=500$. 
The thick line indicates the maxima of fidelity reached by each 
site at different times.}
\label{fig:fid01}
\end{figure}

One can consider a fermion initially 
localized at the time $t = 0$ around the site $j=0$:
\begin{equation}
    \ket{\psi_{0}(0)}=\sum_j g_j c^\dag_{j}\ket{0} \, ,
\end{equation}
with a square wave packet distribution extended over $\lambda=2M+1$ sites:
\begin{equation}
    g_l= \left\{\begin{array}{cc}
      \frac{1}{\sqrt{2M+1}}   & \mbox{      If $ -M \leq l \leq M$} \\
      0                       & \mbox{   elsewhere} \, . \\
    \end{array}
    \right.
\end{equation}
After the state evolution $\ket{\psi_0(t)}=e^{-iHt} \ket{\psi_{0}(0)}$,
the capacity for the channel to produce QST from the site $j = 0$ 
to site $j = d$ can be measured by the square projection of the evolved 
state on the initial state localized on the site $d$:
\begin{equation}
 F_d(t)=  \left | \braket{\psi_{d} (0) |\psi_0(t)} \right|^2 \simeq 
A(t) \, e^{-\frac{
    \left[d+2wt \sin\phi \right]^2}{2 \sigma_F^2(t)}}\,,
\end{equation}
with 
\begin{equation}\label{eqn:amp}
    A(t)=\frac{3 \lambda^2}{\pi\sqrt{(\lambda^2-1)^2+144 w^2 t^2 \cos^2 \phi}} \, \,,
\end{equation}
and 
\begin{equation}\label{eqn:var}
    \sigma_F^2(t)=  \frac{(\lambda^2-1)^2+144 w^2 t^2 \cos^2 \phi}{12 (\lambda^2-1)} \, . 
\end{equation}
For $\phi=-\pi/2$, the dispersion becomes approximately linear and the 
wave-packet 
does not diffuse. Moreover, it propagates with velocity $v=2w$ 
causing an enhancement in the fidelity (see Fig. \ref{fig:fid01}). 
In particular, $F_d(t)$ assumes its maximum 
value at the approximate time
\begin{equation}\label{eqn:tmax2}
   t^\ast = \frac{d}{2 w}.
\end{equation}

\subsection{Violation of the area-law in long-range systems}

The study of the ground-state entanglement properties 
plays an essential role in the characterization of a quantum many-body system. 
In this Section, we show how the introduction of a gauge potential in a free-fermion model 
with a long-range hopping can qualitatively change the scaling behavior of the ground state entanglement.
The amount of entanglement of a pure state is well quantified 
by the so-called Entanglement Entropy (EE), defined as follows. 
Partitioning a given system in 
two subsystems $A$ and its complement $\bar{A}$, the EE is 
the Von Neumann entropy $S$ of one of the two subsystems (say $A$) 
calculated from its reduced density matrix $\rho_A$:
\begin{equation}
S=-Tr\left(\rho_A \ln \rho_A\right)\,.
\end{equation}

Typically for gapped short-range quantum systems 
(where gapped means with a finite energy difference of the first excited level, compared to the ground state energy), 
the EE grows as the boundary of the subsystem $A$ , i.e., 
for a system in $d$ dimensions the EE scales as 
$S \propto \partial L^{d - 1}$. This is commonly known as the {\em area law} 
\cite{eisert10}. 
The physical origin of this law is that entanglement is appreciably nonvanishing only between 
parts of the system very close each others, since the quantum correlation functions between two points decay exponentially with their distance, with 
a finite decay constant $\xi$ that increases as the mass gap decreases.
At variance, for short-range free fermionic systems at a critical (gapless) point
it has been shown that the divergence of $\xi$ (resulting in an algebraic decay of quantum correlations) produces a logarithmic correction of the area law, 
$S \propto L^{d-1} \ln L$ \cite{wolf06,gioev06}, so that in one dimension 
one expects to find $S\propto \ln L$. 
A more relevant, non-logarithmic, violation of the area law is obtained when 
$S\propto L^\beta$ with $d-1<\beta<d$. When $\beta=d$ one has a 
{\em volume law}.

Referring to free fermions on a lattice, in order to find violation to the area law in gapped regimes, 
one has to introduce longer-range connections, changing the Fermi surface in a suitable way. In one-dimensional short-range 
systems, the Fermi surface is typically composed by a finite set of points. 
This is what happens also in the simplest long-range models when, 
despite of the long-range hoppings, strong entanglement is 
created only between closed lattice sites. At variance, if  the Fermi surface 
is a  set of points with finite dimension, 
it can occurs that antipodal sites of the lattice becomes maximally entangled 
(Bell pairs).
As a consequence, a bipartition into two connected complementary 
parts would cut a number of Bell pairs of the order of the volume of the 
smaller subsystem, giving rise to a violation of the area law.
To this aim, a long-range connection appears useful but not sufficient.

A possible way to create a nontrivial Fermi surface is to introduce a 
gauge potential \cite{Gori2015}. Let us consider a model 
with long-range hopping with periodic boundary conditions:
\begin{equation} \label{eqn:htopo}
H=-\sum_j w_{i,j}  c^\dagger_i c_{j} + h.c.
\end{equation}
with 
\begin{equation}
w_{i,j}=w\frac{e^{i \, \phi \, d_{i,j}  }}{|d_{i,j}|^\alpha},
\label{magnetic_phase}
\end{equation}
where $\phi=\frac{2\pi}{N_S} \Phi$, being $\Phi$ a constant, $N_S$ the number 
of sites. The filling $f$ is defined to be $f=N/N_S$, where $N$ is the number 
of sites. 
$d_{i,j}$ is the oriented distance between the sites $i$ and $j$, 
\begin{equation}\label{eqn:distphase}
d_{i,j}=\left\{
\begin{array}{c}
(i-j)  \quad \textrm{if}\:\left|i-j\right|\leq N_{S}-\left|i-j\right|  \\
-N_{S}+\left|i-j\right| \quad \textrm{otherwise}.
\end{array} 
\right.
\end{equation}
Due to the translational invariance, the eigenstates are 
plane waves, and, for finite $N_{S}$,
the spectrum is given by:
\begin{equation}
E_{k}=-2w
\left\{
\begin{array}{cc}
\mbox{\ensuremath{\sum_{m=1}^{\frac{N-1}{2}}\frac{1}{m^{\alpha}}\cos\left((k+\phi)m\right)}} & \textrm{for odd}\: N_{S} \, ,\\
{}\\
\sum_{m=1}^{\frac{N}{2}-1}\frac{1}{j^{\alpha}}\cos\left(\left(k+\phi\right)m\right)+
\frac{\cos\left(\pi n_{k}\right)}{2\left(\frac{N_S}{2}\right)^{\alpha}} & \textrm{for even}\: N_{S} \, .
\end{array}
\right.
\label{lerch}
\end{equation}

\begin{figure}[htbp]
\begin{center}
\includegraphics[height=8cm,angle=-90]{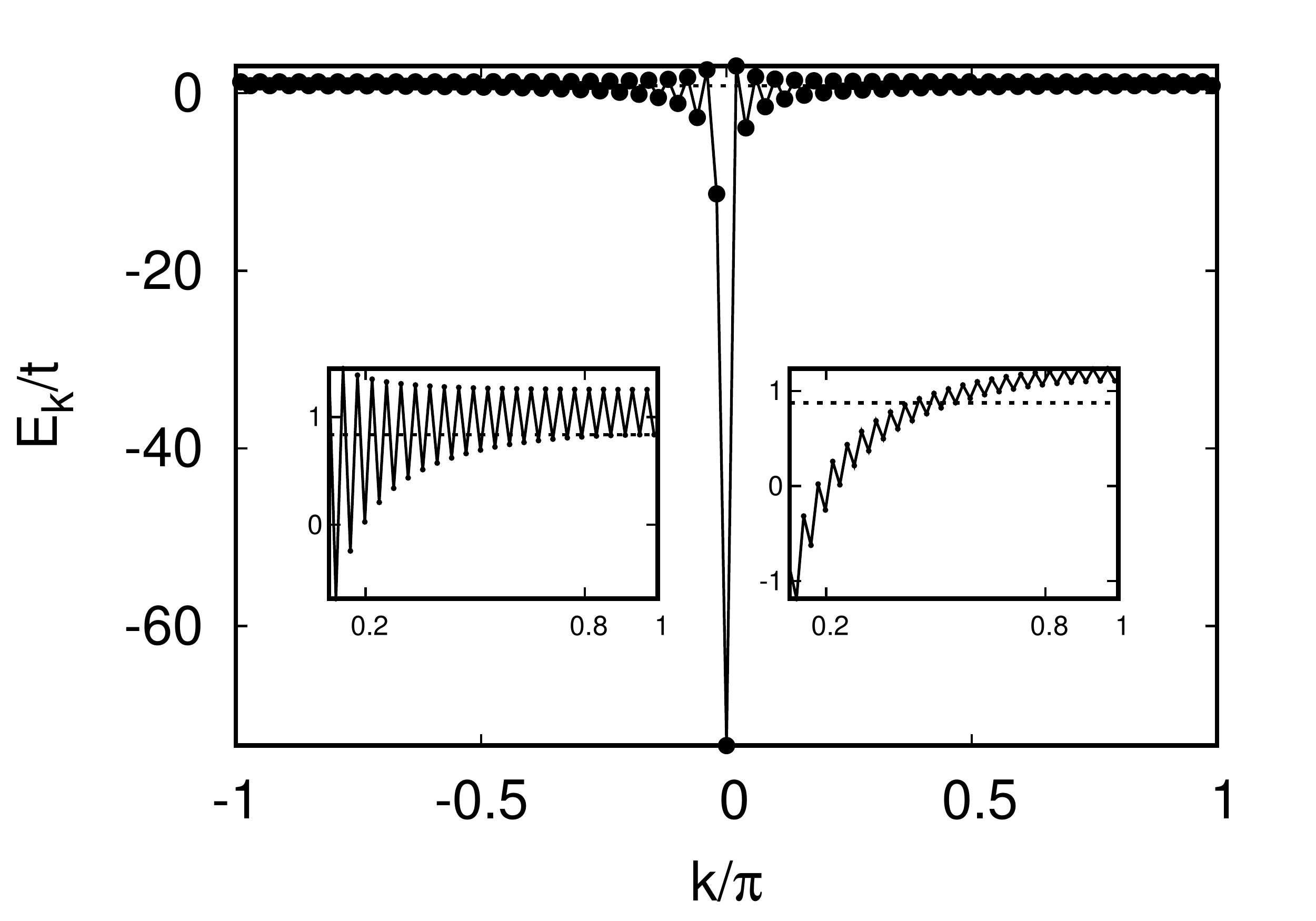}
\end{center}
\caption{Spectrum of the Hamiltonian in Eq. (\ref{eqn:htopo})
with $\Phi=0.1$, $\alpha= 0.1$, filling factor $f=0.5$ and $N_S=100$.
Left inset: detail of the main plot showing the
alternating occupation of the modes $k$, the Fermi energy corresponding 
to the dashed line.
Right inset: decrease of the alternating occupation 
with increasing $\alpha$. We set
$\Phi=0.1$, $\alpha=0.4$, $f=0.5$ and $N_S=100$.}
\label{fig:spectr}
\end{figure}

For $\phi=0$, the single-particle spectrum is always monotonous in the interval 
$k \in [0 ,\pi ]$, while for $\phi\neq0$ the spectrum 
can split in two branches for 
$\alpha<\alpha_c<1$,  where the critical value $\alpha_c$ depends both on $N$ and $\phi$. 
This means that at fixed $\phi$ and $N_S \gg 1$
at half-filling, 
all the momenta $k$ are occupied in an alternating way, 
as shown in Fig. \ref{fig:spectr}. Thus, 
for $\alpha<\alpha_c$ and at half-filling, the ground-state is 
a Bell-paired state, and the EE grows linearly with $N_s$ 
(with slope $\ln{2}$), resulting in a volume law and the Fermi surface 
has a fractal 
counting box dimension  $d_{box}=1$.
On the contrary, when 
$\alpha>\alpha_c$ only a fraction of the momenta are occupied in an 
alternating way, since the ``zig-zag'' structure of the dispersion relation is partially lost. As a result, 
the slope of the EE decreases. 

We conclude by observing that as long as the dispersion 
is such that the half-filling  occupation is alternate in $k$,  
entanglement is created between antipodal sites and the system 
violates the area law behavior, in favour of a volume law.

\end{document}